\documentclass[12pt]{iopart}

\usepackage{amssymb}
\usepackage{graphicx}
\usepackage{bm}
\usepackage{color}
\usepackage{cancel}
\usepackage{slashed}

\newcommand{\Ignore}[1]{}

\newcommand{\Ket}[1]{\left\vert #1\right\rangle}
\newcommand{\Bra}[1]{\left\langle #1\right\vert}

\newcommand{\ii}{\mathrm{i}}

\newcommand{\eqref}[1]{(\ref{#1})}

\begin{document}

%\title{Strong-coupling induced frequency squeezing leading to synchronization in a star network}
\title{Star network synchronization led by strong-coupling induced frequency squeezing}

\author{Benedetto Militello}
\address{Dipartimento di Fisica e Chimica, Universit\`a degli Studi di Palermo, Via Archirafi 36, I-90123 Palermo, Italy}
\address{I.N.F.N. Sezione di Catania, Italy}

\author{Dariusz Chru\'sci\'nski}
\address{Institute of Physics, Faculty of Physics, Astronomy and Informatics Nicolaus Copernicus University, Grudziadzka 5/7, 87-100 Torun, Poland}

\author{Anna Napoli}
\address{Dipartimento di Fisica e Chimica, Universit\`a degli Studi di Palermo, Via Archirafi 36, I-90123 Palermo, Italy}
\address{I.N.F.N. Sezione di Catania, Italy}

\begin{abstract}
We consider a star network consisting of $N$ oscillators coupled to a central one   which in turn is coupled to an infinite set of oscillators (reservoir), which makes it leaking. Two of the $N+1$ normal modes are dissipating, while the remaining $N-1$ lie in a frequency range which is more and more squeezed as the coupling strengths increase, which realizes synchronization of the single parts of the system.
\end{abstract}

\maketitle

\section{Introduction}

The possibility of synchronizing the dynamics of two or more, different or not, physical systems, thus realizing coherent evolutions, has attracted a lot of attention in many different disciplines like physics, chemistry, biology but also social science~\cite{ref:Pikovski2003}. Synchronization phenomenon is relevant  also in view  of applications in neurosciences and
even in medicine~\cite{ref_Strogatz, ref_AngeliniPRE2004}.
It is a rich and intriguing phenomenon that was originally studied in classical systems where it is effectively described by the Kuramoto model ~\cite{ref_AcebronRMP2005}. Detailed studies (based on the laws of classical mechanics) of occurrence of synchronization in classical systems, such as coupled pendula and metronomes, have been reported~\cite{ref_MaiantiAJP2009,ref_PantaleoneAJP2002}.

During the last decades the search of synchronized behaviors has been extended to quantum platforms \cite{ref:Galve2017}.
In this context attempts to adapt the Kuramoto model to the quantum realm have been made \cite{ref_deMendozaPRE2014}. Beyond the Kuramoto model, studies of synchronization in quantum
systems have been developed \cite{ref_ZambriniPRA2010,ref_ZambriniPRA2012,ref_ZambriniSciRep2013}, also with the aim of providing proper definition and measure tools of the degree of synchronization \cite{ref_FazioPRL2013,ref_FazioPRA2015,ref_ArmourPRA2015}.

In the last years, other than the natural archetypical quantum system, i.e. the harmonic oscillator, fundamental class of quantum systems have been considered in order to prove the possibility of realizing synchronization processes. A few uncoupled spins interacting with a common environment~\cite{ref_Plastina2013}  as well as ensembles of dipoles~\cite{ref_Zhu2015}  have been considered. More recently, collective behavior of many spins has been studied in order to establish a connection between synchronization processes and superradiance or subradiance~\cite{ref_Bellomo2017}. A further extension present in the literature is the dynamical alignment of optomechanical systems~\cite{ref_Li2016,ref_Li2017,ref_Vitali2017}, as well as hybrid systems like two-level atoms and oscillators~\cite{ref_ArmourPRA2015, ref_Qiu2015, ref_Boccaletti}. 
Some works tracing back the origin of quantum synchronization to dissipation have appeared \cite{ref_ZambriniSciRep2013, ref:Manzano2013, ref:Militello2017}.

In this paper we try to give a quantum counterpart of typical scenario of synchronization in classical mechanics. It is well known that two or more metronomes lying on a common platform which in turn can move (for example being place above two cans) and dissipate energy to the ground eventually synchronize. We then consider $N$ quantum harmonic oscillators (the metronomes counterparts) interacting with another oscillator (corresponding to the platform) which is coupled to a an environment consisting of an infinite set of harmonic oscillators (corresponding to the ground).

We will show that in the strong coupling limit (when the coupling of each oscillator with the leaking one is big) we obtain a twofold effect: on the one hand, two leaking normal modes appear, and on the other hand the remaining non-leaking modes occupy a frequency range whose amplitude becomes smaller and smaller when the strength of the coupling with the leaking mode increases. Our analysis thus allows to bring into light in a very clear way the origin of the mechanism that, at least for the model envisaged in the paper, leads to synchronization phenomenon.

\section{The Model}

Consider a system governed by the following Hamiltonian:
%\begin{subequations}
\begin{eqnarray}\label{eq:appHtot}
  H &=& H_S + H_B + H_I\,,
\end{eqnarray}
with the system -- a star network-- Hamiltonian,
\begin{eqnarray}\label{eq:appHSN}
  \nonumber
  H_S &=& \frac{1}{2} \sum_{i=1}^N \left( \frac{p_i^2}{m} + k_i x_i^2 \right) + \frac{1}{2} \left( \frac{p_{N+1}^2}{m} + k_{N+1} x_{N+1}^2  \right)\\
      &+& \frac{1}{2}\sum_{i=1}^N g_{i} (x_{N+1}-x_i)^2\,,
\end{eqnarray}
the Hamiltonian of the bath (reservoir),
\begin{eqnarray}\label{eq:appHB}
  H_B &=& \sum_j \frac{p_j^2}{2m_j}+\frac{1}{2} k_j^{(b)}y_j^2\,,
\end{eqnarray}
and the interaction Hamiltonian given by
\begin{eqnarray}\label{eq:appHI}
  H_I &=& x_{N+1} \, \sum_j \gamma_j y_j\,,
\end{eqnarray}
where $y_j$'s are the coordinates referred to the bath
oscillators.
%\end{subequations}

There is a single mode of the main system --- the one corresponding to $x_{N+1}$ --- which is a leaking one. We assume that all the masses of the $N+1$ oscillators re equal, a condition that we can always realize through a suitable canonical transformation. In general, the coupling constants $g_i$'s can be positive or negative, in the latter case describing repulsive interactions. In our specific case, since to obtain synchronization we will consider the large $g_i$ limit, in order to prevent instability of the system we will assume $g_i\ge 0$ $\forall i$.

\section{Normal Modes Analysis}

After ordering the $N+1$ coordinates as $(x_1, \ldots , x_N, x_{N+1})$, the whole potential in \eqref{eq:appHSN} can be considered as a quadratic form $\frac 12 \sum_{i,j=1}^{N+1} \mathbf{V}_{ij} x_i x_j$ associated to the following matrix:
%\begin{subequations}
\begin{eqnarray}
\nonumber \mathbf{V}=\left(
\begin{array}{cccccc}
k_1+g_1 & 0 & & \cdots & & -g_1 \\
0 & k_2+g_2 & & \cdots & & -g_2 \\
0 & 0 & & \ddots & & \vdots \\
0 & 0 & & & k_N+g_N & -g_N \\
-g_1 & -g_2 & & \cdots & -g_N & k_{N+1}+N g_\mathrm{av} \\
\end{array}
\right) \,, \\ %
\end{eqnarray}
with the averaged coupling constant
\begin{eqnarray}\label{eq:appgav}
  g_{\mathrm{av}} := \frac 1N \sum_{j=1}^N g_j \,.
\end{eqnarray}
%\end{subequations}
This matrix can be reorganized as:
%\begin{subequations}
\begin{eqnarray}
%\nonumber
\mathbf{V} = (k_\mathrm{av} + g_\mathrm{av}) \mathbf{I} + \mathbf{G} + \mathbf{D} \,,
\end{eqnarray}
with the diagonal matrix
\begin{eqnarray}
\nonumber
\mathbf{D} = \left(
\begin{array}{cccccc}
\delta k_{1} + \delta g_{1} & 0 & & \cdots & & 0 \\
0 & \delta k_{2} + \delta g_{2} & & \cdots & & 0 \\
0 & 0 & & \ddots & & \vdots \\
0 & 0 & & & \delta k_{N} + \delta g_{N} & 0 \\
0 & 0 & & \cdots & 0 & 0 \\
\end{array}
\right)\,, \\ %
\end{eqnarray}
and
\begin{eqnarray}
%\nonumber
\mathbf{G}=\left(
\begin{array}{cccccc}
0 & 0 & & \cdots & & -g_1 \\
0 & 0 & & \cdots & & -g_2 \\
0 & 0 & & \ddots & & \vdots \\
0 & 0 & & & 0 & -g_N \\
-g_1 & -g_2 & & \cdots & -g_N & \Delta \\
\end{array}
\right) \, . % \\ %
\end{eqnarray}
where we have introduced the averaged Hook's constant,
\begin{eqnarray}\label{eq:appkav}
  k_\mathrm{av} := \frac{1}{N+1} \sum_{j=1}^{N+1} k_j \,, %\qquad
\end{eqnarray}
and
\begin{equation}\label{}
  \delta k_j \equiv k_j-k_\mathrm{av}\,, \ \  \delta g_i = g_i -  g_\mathrm{av}\, ,
\end{equation}
together with
\begin{equation}
\Delta = \delta k_{N+1} + (N-1) g_\mathrm{av}\,.
\end{equation}
%\end{subequations}
In what follows we consider the following regime:
\begin{equation}
  |\delta g_j|, |\delta k_j| \ll g_\mathrm{min}\,,
\end{equation}
with $g_\mathrm{min}=\min_j g_j$, one may treat $\mathbf{D}$ matrix as a perturbation to $(k_\mathrm{av} + g_\mathrm{av}) \mathbf{I} + \mathbf{G}$, that is, to diagonalize $\mathbf{V}$ one diagonalizes $\mathbf{G}$ and then looks for perturbative corrections induced by $\mathbf{D}$.

\subsection{Diagonalization of $\mathbf{G}$}

 The eigenvalues of
$\mathbf{G}$ are: $G_0=0$ ($N-1$ eigenstates), and two singlets
\begin{equation}\label{}
G_\pm= \frac 12 (\Delta\pm\sqrt{\Delta^2+4\Lambda^2})
\end{equation}
with $\Lambda^2 = \sum_j g_j^2$. The corresponding \lq eigenvectors\rq\, are:
%\begin{subequations}
\begin{equation}\label{eq:plusminusmodes}
  \bar{x}_\pm =  \sum_{j=1}^{N} \alpha_j^{\pm} x_j + \alpha_{N+1}^{\pm} x_{N+1}\,,
\end{equation}
with
\begin{equation}
  \alpha_j^{\pm} / \alpha_{N+1}^{\pm} = - g_j / G_\pm \,, \qquad j=1,..., N
\end{equation}
and for the degenerate $(N-1)$--dimensional subspace we can take
the following orthonormal set:
%\begin{widetext}
\begin{eqnarray}\label{eq:zeromodes}
\nonumber
\begin{array}{lcl}
\bar{x}_{0,1} &=& \cos\theta_1 x_1 + \sin\theta_1 x_2\,,  \\
\bar{x}_{0,2} &=&  \cos\theta_2(-\sin\theta_1 x_1 + \cos\theta_1
x_2) + \sin\theta_2 x_3\,, \\
\bar{x}_{0,3} &=&  \cos\theta_3(-\sin\theta_2(-\sin\theta_1 x_1 +
\cos\theta_1
x_2) + \cos\theta_2 x_3) + \sin\theta_3 x_4\,,  \\
\vdots & \\
\end{array}\\
\end{eqnarray}
with
\begin{eqnarray}
\nonumber
\begin{array}{lcl}
\tan\theta_1 &=& -g_1/g_2 \\
\tan\theta_2 &=& -\sqrt{g_1^2+g_2^2}/g_3\\
\tan\theta_3 &=& -\sqrt{g_1^2+g_2^2+g_3^2}/g_4\\
\vdots & \\
\end{array}\\
\end{eqnarray}
%\end{subequations}
%\end{widetext}
The $0$-modes do not involve the coordinate $x_{N+1}$ and then are not subjected to decay processes, therefore we will call them \lq protected\rq\, modes.

\subsection{Perturbation treatment of $\mathbf{D}$}

In order to complete our analysis we need to treat $\mathbf{D}$ perturbatively.

First of all, we evaluate the eigenvalue zeroth-order  correction related to the normal modes $\tilde{x}_\pm$, which is easily done by evaluating the relevant diagonal matrix elements:
\begin{eqnarray}\label{eq:appDExpValPM}
  \nonumber %
   \Delta k_\pm &\equiv& (\bar{x}_\pm, \mathbf{D} \bar{x}_\pm) \\
   &=& (\alpha_{N+1}^\pm)^2 \Delta + \sum_{j=1}^{N} (\alpha_j^\pm)^2 (\delta k_j + \delta g_j)\,.
\end{eqnarray}
Then we should proceed by evaluating the first order correction to the \lq eigenvectors\rq\,.
Such correction are of the order $(\delta k_j + \delta g_j) / g_\mathrm{av}$, so that we can write:
\begin{eqnarray}\label{eq:AlmostBarTilde}
\tilde{x}_{\pm} = \bar{x}_{\pm} + o(\xi)\,,
\end{eqnarray}
where
\begin{eqnarray}
\xi = \max_j[|\delta k_j| + |\delta g_j|] / g_\mathrm{av}\,.
\end{eqnarray}

For the protected modes we need to diagonalize the restriction of $\mathbf{D}$ in the relevant eigenspace. Following this procedure, we will obtain  a correction to the \lq eigenvectors\rq\,:
\begin{eqnarray}
\{\bar{x}_{0,j}\} \rightarrow  \{\bar{z}_{0,j} = \sum_k
O_{jk}\bar{x}_{0,k}\}
\end{eqnarray}
%\RED{I propose to change $\{\bar{\bar{x}}_{0,j}\}$ by
%$\{{\bar{y}}_{0,j}\}$
%\begin{eqnarray}
%\{\bar{x}_{0,j}\} \rightarrow \{ \bar{y}_{0,j} = \sum_i O_{ji} \bar{x}_{0,i} \}
%\end{eqnarray}
%with orthogonal matrix $O_{ij}$ },
and a correction to the eigenvalues, $\{\Delta k_{0, j}\}$. Then
the eigenvectors must be corrected to the first order, then
obtaining:
\begin{eqnarray}
\tilde{x}_{0,j} = \bar{z}_{0,j} + o(\xi)\,.
\end{eqnarray}
The corrected Hook's constants are then given by:
%\begin{subequations}
\begin{equation}\label{eq:HookConstants}
  k_\pm = g_\mathrm{av} + k_\mathrm{av} + G_\pm + \Delta k_\pm\,,
\end{equation}
and, for the protected modes:
\begin{equation}
  k_{0,j} = g_\mathrm{av} + k_\mathrm{av} + \Delta k_{0,j}\,.
\end{equation}
%\end{subequations}

A very special case is that of two oscillators. Indeed, if $N=2$ then the $0$-mode subspace is a singlet (then $\bar{z}_{0,j}=\bar{x}_{0,j}$) and the correction of the eigenvalues requires only the evaluation of the relevant diagonal matrix element of $\mathbf{D}$:
\begin{eqnarray}\label{eq:appDExpVal0}
  \nonumber %
  \Delta k_{0,1} &\equiv& (\bar{x}_{0,1}, \mathbf{D} \bar{x}_{0,1}) \\
  \nonumber %
  &=& \cos^2\theta_1 (\delta k_1 + \delta g_1) + \sin^2\theta_1 (\delta k_2 + \delta
  g_2)\\
  %\nonumber %
  &=& \frac{ g_2^2 (\delta k_1 + \delta g_1) + g_1^2 (\delta k_2 + \delta
  g_2) }{g_1^2 + g_2^2}
  \,,
\end{eqnarray}
\begin{eqnarray}\label{eq:appDExpVal0MOde}
  \tilde{x}_{0,1} = \bar{x}_{0,1} + o(\xi)\,.
\end{eqnarray}

\section{Uniform vs Almost Uniform Model}

Let us consider the case of $N$ different oscillators (that is different $k_i$'s) but with the same coupling strengths to the leaking mode ($g_i=g$, $\forall i$), which means $\delta g_i = 0$, $\forall i$.

Now, whatever it is the specific value of $g$, the structure of the \lq eigenvectors\rq\, in \eqref{eq:plusminusmodes} and \eqref{eq:zeromodes} does not change, as well as the matrix $\mathbf{D}$ is not modified.

%On the contrary, all the matrix elements of $\mathbf{G}$ but
%$\Delta$ are proportional to $g$ and then its eigenvalues are
%$g$-dependent. Moreover, in the limit $g \gg \delta k_j$, which
%implies $\Delta \approx (N-1)g$, we find that $G_\pm$
%are almost proportional to $g$. %: $G_\pm(g)\approx g G_\pm(1)$.
Therefore, by making a first order Taylor expansion in the expression of the frequencies of the protected modes, we get:
\begin{equation}\label{eq:appFreqSqueezing}
\omega_{0,j} \equiv \sqrt{\frac{k_{0,j}}{m}} \approx
\sqrt{\frac{g+k_\mathrm{av}}{m}} \, \left[1+\frac{\Delta
k_{0,j}}{2(g+k_\mathrm{av})}\right] \,.
\end{equation}

Let us define: $\Delta k = \max |\Delta k_{0,j}|$, $\delta k =
\max_i |\delta k_i|$, which are of the same order (see for example
the special case in \eqref{eq:appDExpVal0} and put $\delta
g_1=0$). We can then say that $\Delta\omega$ (the amplitude of the
frequency range where all the corrected frequencies of the
$0$-mode multiplet lie) is of the order of $\Delta
k/\sqrt{m(g+k_\mathrm{av})}$. This quantity is smaller than the
original frequency range, which is approximately given by $\delta
k/\sqrt{m k_\mathrm{av}}$. The higher $g$, the tighter is the
frequency distribution of the \lq preserved\rq\, modes.

This squeezing of the frequency range is probably the main result
of this paper. Indeed, the decay of the leaking normal modes is
not enough to justify the occurrence of synchronization. It is
also necessary that the surviving modes are characterized by the
same frequency. This essentially happens when $g$ is assumed to be
much larger than all $\delta k_j$, for the uniform model.

The prediction of a frequency squeezing, though obtained for the
uniform model, can be easily obtained for a non-uniform model
provided it satisfies the condition that all $\delta g_i$ are kept
smaller than a certain quantity, say $\delta g$, in spite of the
fact that each $g_i$ can increase (this can be easily obtained for
example when the $g_i$'s are equally lifted: $g_i\rightarrow
\xi+g_i$), so that in the limit $g_i\rightarrow \infty$ one has
$\delta g_i / g_j \rightarrow 0$. Under such hypothesis, it is
evident that the frequency squeezing in
\eqref{eq:appFreqSqueezing} still holds.

\section{Evolution of Quantum States}

To complete our analysis we want to apply the previous theory to
the study of an evolution.

\subsection{Canonical transformation}

Consider the following canonical transformations of the $N+1$ coordinates and momenta
%\begin{subequations}
\begin{eqnarray}
  \label{eq:TransformX}
  \tilde{x}_j = \sum_l b_l^{(j)} x_l \, \Longleftrightarrow \, x_l = \sum_j \epsilon_j^{(l)} \tilde{x}_j \,, \\
  \label{eq:TransformP}
  \tilde{p}_j = \sum_l c_l^{(j)} p_l \, \Longleftrightarrow \, p_l = \sum_j \eta_j^{(l)}
  \tilde{p}_j \,,
\end{eqnarray}
with $j=1,... N+1$ and $l=\pm, "0,n"$, where the coefficients
should satisfy the following conditions to preserve the
commutation relations:
\begin{eqnarray}
\sum_l b_l^{(j)} c_l^{(j')} &=& \delta_{jj'}\,,\qquad\forall j,j'\,, \\
\sum_l \eta_l^{(j)} \epsilon_l^{(j')} &=&
\delta_{jj'}\,,\qquad\forall j,j'\,.
\end{eqnarray}
%\begin{eqnarray}
%  \sum_l c_l^{(j)} d_l^{(j)} = 1 \,,\qquad \forall j\,.
%\end{eqnarray}
%\end{subequations}

Consequently, the annihilation operators are transformed as
follows:
\begin{eqnarray}
  \nonumber
  a_j &=& \sum_k \left( \epsilon_k^{(j)} \sqrt{\frac{\omega_j}{\tilde{\omega}_k}} + \eta_k^{(j)} \sqrt{\frac{\tilde{\omega}_k}{\omega_j}} \right) \tilde{a}_k  \\
  &+& \sum_k \left( \epsilon_k^{(j)} \sqrt{\frac{\omega_j}{\tilde{\omega}_k}} - \eta_k^{(j)}
  \sqrt{\frac{\tilde{\omega}_k}{\omega_j}}
  \right)\tilde{a}_k^\dag\,,
\end{eqnarray}
and
\begin{eqnarray}
  \nonumber
  \tilde{a}_j &=& \sum_k \left( b_k^{(j)} \sqrt{\frac{\tilde{\omega}_j}{\omega_k}} + c_k^{(j)} \sqrt{\frac{\omega_k}{\tilde{\omega}_j}} \right) a_k  \\
  &+& \sum_k \left( b_k^{(j)} \sqrt{\frac{\tilde{\omega}_j}{\omega_k}} - c_k^{(j)}
  \sqrt{\frac{\omega_k}{\tilde{\omega}_j}}
  \right)a_k^\dag\, ,
\end{eqnarray}
where $\tilde{\omega}_j$ denotes the corresponding frequencies of
the eigenmodes.

\subsection{Markovian evolution of the star network}

To derive the Markovian master equation for the evolution of the
system one may follow the standard approach assuming for example
weak coupling between system and the bath and assume that
initially the bath was in the thermal state at the temperature
$T$. Since the 0-modes (i.e., the protected modes) involve the
modes $\bar{x}_\pm$ to the order $\xi \ll 1$, deriving the master
equation in the normal mode representation one finds that the
0-modes are characterized by decay rates which are of the order
$\xi$ with respect to the decay rates related to the $\pm$ modes.
Therefore, with a good degree of approximation, the complete
evolution of the system may be evaluated through the following
Markovian master equation:
%\begin{subequations}
\begin{eqnarray}
  \nonumber
  \dot\rho &=& {\cal L}_0 \rho + {\cal L}_1 \rho + \xi {\cal D}_0 \rho \,\\
  &\approx&  {\cal L}_0 \rho + {\cal L}_1 \rho \,,
\end{eqnarray}
where
\begin{eqnarray}
  {\cal L}_0 \rho &=& -\ii [H_S^{0}, \rho]\,,\\
  {\cal L}_1 \rho &=& -\ii [H_S^{1}, \rho] + {\mathcal D}_+ \rho + {\mathcal D}_-\rho\,,
\end{eqnarray}
with the corresponding Hamiltonians:
\begin{eqnarray}
  H_S^{0} &=& \sum_j\tilde{\omega}_{0,j}\tilde{a}_{0,j}^\dag\tilde{a}_{0,j}\,,\\
  H_S^{1} &=& \tilde{\omega}_{+}\tilde{a}_{+}^\dag\tilde{a}_{+} + \tilde{\omega}_{-}\tilde{a}_{-}^\dag\tilde{a}_{-}\,,
\end{eqnarray}
and dissipators in the dissipative sectors:
\begin{eqnarray}
  \nonumber
  {\mathcal D}_\pm \rho &=& \gamma_\pm (N(\tilde{\omega}_\pm)+1) \left[\tilde{a}_\pm \rho \tilde{a}_\pm^\dag - \frac{1}{2}\left\{ \tilde{a}_{\pm}^\dag\tilde{a}_{\pm}, \rho
  \right\}\right] \\
  % \nonumber
  &+& \gamma_\pm N(\tilde{\omega}_\pm) \left[\tilde{a}_\pm^\dag \rho \tilde{a}_\pm - \frac{1}{2}\left\{ \tilde{a}_{\pm}\tilde{a}_{\pm}^\dag, \rho
  \right\}\right] \, .
\end{eqnarray}
%\end{subequations}
Finally, ${\cal D}_0$ is a dissipator describing the decay of the
0-modes, which we neglect being of $\xi$ order.

To better understand the origin of this microscopic master equation, consider the Hamiltonian in Eq.(\ref{eq:appHI}) which is responsible for the decay of the $x_N$ mode. Once the variable $x_N$ is expressed as a linear combination of the normal mode coordinates % $x_N =\beta_+ \tilde{x}_+ + \beta_- \tilde{x}_- + o(\xi)$,  with $\beta_\pm = -G_\pm / \sqrt{G_+^2 - G_-^2}$ 
--- $x_N =\cos\theta\, \tilde{x}_+ + \sin\theta\, \tilde{x}_- + o(\xi)$ with $\tan\theta = -G_-/G_+$, coming from Eqs. (\ref{eq:plusminusmodes}) and (\ref{eq:AlmostBarTilde}) --- 
we can see that the dissipators associated to the modes $\tilde{x}_+$ and $\tilde{x}_-$ naturally emerge form the standard derivation of the microscopic Markovian master equation of a damped harmonic oscillator \cite{ref:Petru, ref:Gardiner}.  At the same time, the negligibility of the dissipators related to the $0$-modes is well visible. In fact, we have $N+1$ uncoupled oscillators, each one interacting with the environment, two of them with significant strengths, $N-1$ with negligible strengths.  
It is worth noting that, generally speaking, since the modes $\tilde{x}_+$ and $\tilde{x}_-$ interact with the same bath, cross terms could appear in the dissipator. However, since we are deriving the master equation in the standard Born-Markov approximation, which requires also the secular approximation, and since the two oscillators have quite different frequencies (see eq.(\ref{eq:HookConstants})), such cross terms disappear. Finally, we again underline that our master equation is correct up to terms of the order $\xi$, because this is the precision of our derivation of the normal modes.

The evolution evaluated on the basis of the previous master equation can be factorized as follows:
\begin{eqnarray}
  \rho(0) \rightarrow \rho(t) = {\cal S}(t) \rho(0) = {\cal S}_0(t) {\cal S}_1(t)
  \rho(0)\,,
\end{eqnarray}
with ${\cal S}_0(t)$ and ${\cal S}_1(t)$ generated by ${\cal
L}_0$ and  ${\cal L}_1$, respectively: ${\cal S}_j=\exp({\cal
L}_jt)$, with $j=0,1$.

Suppose now that the system is prepared in a pure quantum state

$$\Ket{\psi(0)} = \sum c_{\{n_{0,j}\}\{n_\pm\}}
\Ket{\{n_{0,j}\}}\Ket{\{n_\pm\}}\,.$$
 Here $\Ket{\{n_{0,j}\}}\Ket{\{n_\pm\}}$ denotes a Fock state of the
$N+1$ normal modes, and in particular $\Ket{\{n_{0,j}\}}$ is a
Fock state of the protected modes, while $\Ket{\{n_\pm\}}$ is a
Fock state of the leaking modes. The relevant density operator
reads

%\begin{widetext}
\begin{eqnarray}
  %\nonumber
  \rho(0) = \Ket{\psi(0)}\Bra{\psi(0)}
  %\nonumber
= \sum c_{\{n_{0,j}\}\{n_\pm\}} c_{\{n_{0,j}'\}\{n_\pm'\}}^*
  \Ket{\{n_{0,j}\}}\Ket{\{n_\pm\}}\Bra{\{n_{0,j}'\}}\Bra{\{n_\pm'\}}\,. % \\
\end{eqnarray}
%\end{widetext}
Under the action of ${\cal S}_1(t)$ for a sufficient large time
all the terms with $\{n_\pm\}\not=\{n_\pm'\}$ go to zero (all
coherences are destroyed by the dissipative evolution) while the whole set of
the terms with $\{n_\pm\}=\{n_\pm'\}$ is eventually mapped to the thermal
state of the two leaking modes $\rho_\pm^{(T)}$. Therefore, after
a sufficient long time one has:
%\begin{widetext}
\begin{eqnarray}
  \nonumber
  {\cal S}(t)\rho(0) &\approx&
  \left(\sum c_{\{n_{0,j}\}\{n_\pm\}} c_{\{n_{0,j}'\}\{n_\pm\}}^*
  {\cal S}_0(t)\Ket{\{n_{0,j}\}}\Bra{\{n_{0,j}'\}}\right)\otimes \rho_\pm^{(T)}  \\
  &=& \left(\sum c_{\{n_{0,j}\}\{n_\pm\}} c_{\{n_{0,j}'\}\{n_\pm\}}^*
  U_0(t)\Ket{\{n_{0,j}\}}\Bra{\{n_{0,j}'\}}U_0(t)^\dag\right)\otimes \rho_\pm^{(T)}
\end{eqnarray}
with $U_0(t)=\exp(-\ii H_S^{0} t)$.
%\end{widetext}

%\begin{widetext}

In particular for $N=2$ one has one only a single protected mode
($\Ket{n_0}$ indicates its generic Fock state) and hence:
\begin{eqnarray}
  {\cal S}(t)\rho(0)  \approx
   U_0(t) \, (\sum_{n_0,n_0'} \alpha_{n_0n_0'} \Ket{n_0}\Bra{n_0'})\, U^\dagger_0(t) \otimes \rho_\pm^{(T)}
\end{eqnarray}
where
\begin{equation}\label{}
  \alpha_{n_0n_0'} = c_{n_0,n_+}c^*_{n_0',n_+} +  c_{n_0,n_-}c^*_{n_0',n_-} \, .
\end{equation}
%\end{widetext}

\subsection{Expectation Values}

After a long time the expectation value of the position
operator of any of the protected modes is
$\langle\tilde{x}_j(t)\rangle = \langle\tilde{x}_j(0)\rangle
\cos(\tilde{\omega}_j t)$, where $\langle\tilde{x}_j(0)\rangle = \sqrt{m\tilde{\omega}_j/(2\hbar)} \Bra{\psi(0)} (\tilde{a}_j + \tilde{a}_j^\dag) \Ket{\psi(0)}$.

Taking into account of the
transformation laws in \eqref{eq:TransformX} one easily finds:
\begin{eqnarray}
  \langle x_l(t)\rangle = \sum_j \epsilon_j^{(l)} \langle\tilde{x}_j(0)\rangle
\cos(\tilde{\omega}_j t) \,.
\end{eqnarray}

A similar result holds for $\langle p_l(t)\rangle$.

It is worth recalling that, at the order we have developed our
analysis, the protected modes do not involve the two leaking modes
in their transformation. Moreover, the frequencies of the
protected modes are very close, in the strong coupling limit
($g_\mathrm{av}\gg \delta g_i, \delta k_i$). The difference
between any couple of such frequencies is of the order of
$\max(\delta g, \delta k) / \sqrt{k_\mathrm{av}+g_\mathrm{av}}$
and then becomes smaller and smaller as $g_\mathrm{av}$ increases.
This makes both $\langle x_l(t)\rangle$ and $\langle
p_l(t)\rangle$ essentially oscillate at a given frequency which is
$\bar\omega = \sqrt{(k_\mathrm{av}+g_\mathrm{av})/m}$.

Again, after a long time, the expectation values of the two
leaking normal modes are zero
($\langle\tilde{x}_+\rangle=\langle\tilde{x}_-\rangle=0$), then we
can get information about the mean values of $x_{N+1}$ and
$\sum_{j=1}^N g_j x_j$.

%\section{A simple example ($N=3$)}

\begin{figure}
%\centering
\includegraphics[width=0.9\textwidth, angle=0]{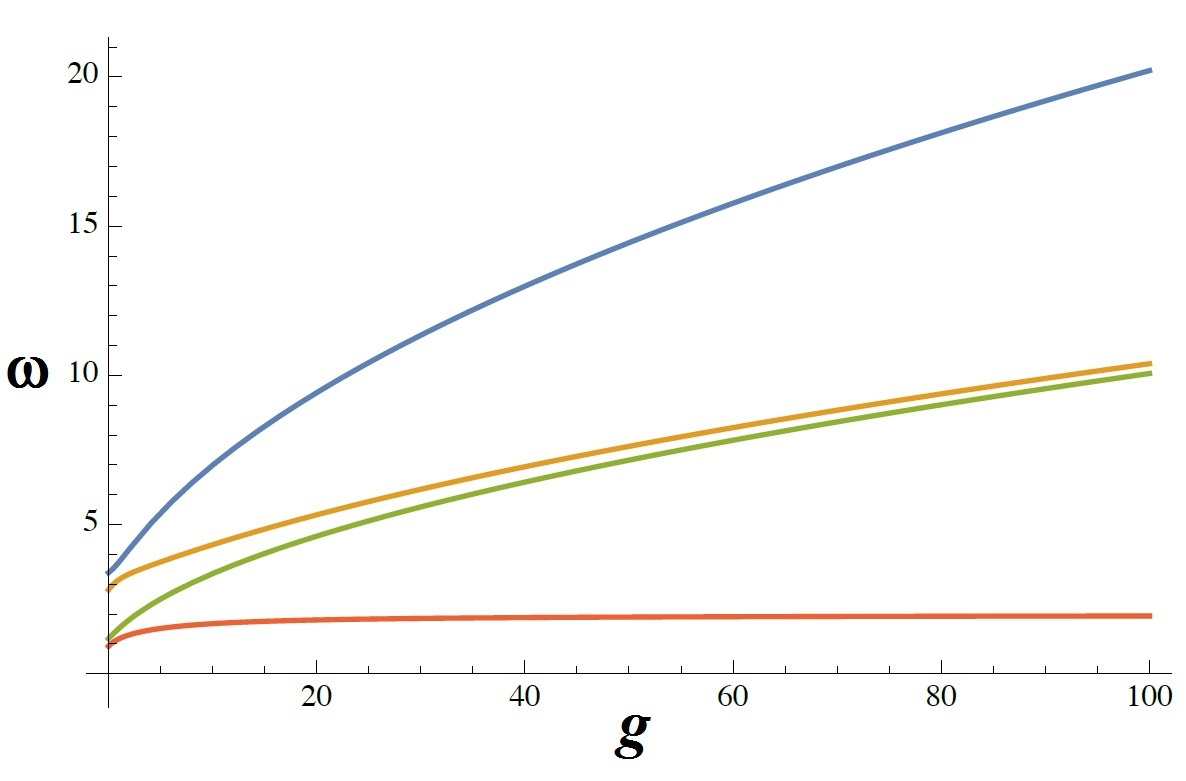} %
\caption{(Color online). Frequencies of the normal modes vs the
average coupling strength $g=g_{\mathrm{av}}$ (in units of $k_1$).
Here $k_2/k_1=0.2$, $k_3/k_1=10$, $g_1/k_1=g/k_1+0.9$,
$g_2/k_1=g/k_1+1$, $g_1/k_1=g/k_1+1.1$. Blue and brown lines
correspond to the frequencies of the decaying modes, while green
and orange lines correspond to the frequencies of the protected
modes.} \label{fig:FreqSq}
\end{figure}

In order to demonstrate the synchronization and the frequency
squeezing in a very simple case, let us consider for example the
simple case of three oscillators all coupled to a fourth one.
According to the previous analysis, we obtain, as shown in figure
\ref{fig:FreqSq} that  the frequencies of the preserved modes get
closer and closer as the coupling constants with the fourth
oscillator increase.

%{\bf [work in progress]}

\section{Conclusions}

In this paper we have considered a system composed by $N$ oscillators coupled to a central one which in turn
interacts with an infinite set of oscillators (reservoir).  This system can be seen as a quantum counterpart of the classical system consisting two or more than two metronomes lying in the same platform. We have shown that, under appropriate conditions, it is possible to foresee synchronization phenomena in the dynamics of the system. In particular the analysis developed put clearly into light how the original intrinsic frequencies of the oscillators are modified by the interaction with the central one leading to a common effective frequency. More in detail we have demonstrated that
two of the $N+1$ normal modes of the system are dissipating modes,
while the remaining $N-1$ lie in a frequency range which is more
and more squeezed as the coupling strength increases. It is just the frequency squeezing phenomenon that allows to the $N$ oscillators
to evolve by swinging in unison. It is worth mentioning that, though also in the case of two oscillators coupled to a third one the system reaches a synchronized regime because of the presence of a single stable mode, the phenomenon of frequency squeezing is visible only with more than two oscillators coupled to the leaking one.

\end{document}